\documentstyle[aps,prl]{revtex}
\begin{document}
\draft

\twocolumn[\hsize\textwidth\columnwidth\hsize\csname @twocolumnfalse\endcsname

\title{Flux-flow in $d$-wave superconductors:
Low temperature universality and scaling}
\author{N.B. Kopnin$^{1,2}$ and G.E. Volovik$^{1,3}$}
\address{
$^1$Landau Institute for Theoretical Physics, 117334 Moscow,
Russia\\
$^2$Laboratoire de Physique des Solides,
Universit\'e Paris-Sud, B{\^a}t 510, 91405 Orsay, France\\
$^3$Helsinki University of Technology, Low Temperature Laboratory,
P. O. Box 2200, FIN-02015 HUT, Finland }

\date{\today}
\maketitle

\begin{abstract}
We demonstrate that superclean $d$-wave superconductors
with $\Delta ^2_\infty \tau /E_F \gg 1$ display a novel type of vortex dynamics:
At low temperatures, both dissipative and transverse components of the flux-flow
conductivity are found to approach universal values even in the limit of
infinite relaxation time. A finite dissipation in the superclean limit is
explained in terms of the Landau damping on zero-frequency vortex modes which
appear due to  minigap nodes in the bound-state spectrum in the vortex core.
In the moderately clean regime the scaling law at low $T$ and low field
is obtained.
\end{abstract}

\


\

]
\narrowtext
The Landau description of low-temperature thermodynamics and low-frequency
dynamics of Fermi and Bose liquids  is based on the concept of
predominating low-energy quasiparticles. In the same way,
the low-frequency dynamics of vortices in Fermi superfluids and  superconductors
appears to be also determined by low-energy excitations.
For Fermi superfluids and
superconductors with a gap in the spectrum, the relevant low-energy
quasiparticles are concentrated in vortex cores. The anomalous spectral branch
of quasiparticles localized in the core, which crosses zero as a
function of the orbital quantum number, was first found by
Caroli, de Gennes and Matricon
\cite{Caroli} for the simplest case of axisymmetric vortices in
$s$-wave superconductors. Effects of these quasiparticles on the vortex
dynamics were considered in a number of papers (see, for example,
\cite{KopKr,KL}).

The $d$-wave superconductivity adds new aspects to this problem:
due to the gap nodes, gapless quasiparticles exist also outside vortex cores.
The low-energy fermions which
are far from vortex cores but still are under the influence of the
vortex superflow velocity (i) introduce non-analytical terms in the free
energy of the mixed state of a
$d$-wave superconductor \cite{d-waveVortex};
(ii) result in a singularity in the vortex density of
states\cite{KopninVolovik} and (iii) lead to a peculiar scaling law for
thermodynamic and kinetic properties of $d$-wave superconductors at low
$T\ll T_c$  in small magnetic fields $B\ll
H_{c2}$ \cite{KopninVolovik,SimonLee,VortexEntropy}.  Here we show
that the vortex dynamics is also influenced by these far-distant excitations
bound to vortices. This results in a new scaling law for
the Ohmic and Hall conductivities in the mixed state of a
$d$-wave superconductor. Moreover, a novel effect appears: the vortex
motion is accompanied by the Landau damping on vortex
modes giving rise to a finite dissipation even in the superclean limit.

It is known \cite{GorkovKalugin,Lee} that, in a $d$-wave superconductor,
impurity
scattering broadens the gap nodes to the impurity band whose width is
$\delta \sim T_c \exp (-\pi \Delta _0\tau )$ in the Born approximation.
For temperatures below the impurity bandwidth $\delta$,
the current response function in the
Meissner state,  ${\bf j}=-Q(\omega ){\bf A}(\omega )$ displays
unusual behavior: The ``conductivity'' $\sigma (0)$ defined as
$Q(\omega \to 0)=n_se^2/mc -i(\omega /c)\sigma (0)$,
saturates to a scattering-independent universal value \cite{Lee,Maki}
corresponding to the Drude conductivity with the effective relaxation time
$1/\tau \sim T_c$. In the present Letter, we demonstrate that a similar
effect exists in the mixed state of $d$-wave superconductors:
Both longitudinal (Ohmic) and transverse (Hall) components of the conductivity
tensor approach finite universal values even in the limit of infinite
relaxation time, it  temperatures and magnetic fields are low enough. This
universal behavior of the flux-flow conductivity is realized in superclean
systems with the mean free path $\ell \gg (E_F/T_c) \xi$. We trace the finite
dissipation in the superclean limit back to the resonant absorption (Landau
damping) on zero-frequency vortex modes which appear due to the nodes in the
minigap
$\omega _0$, i.e., in the distance between bound-state levels in the
vortex core. Note that the width of the abovementioned
impurity band $\delta$ is vanishingly small for superconductors with an
impurity scattering rate satisfying the superclean condition.

We start with the spectrum of relevant low-energy fermionic
quasiparticles assuming that  $T \ll T_c$.
Since our results are not sensitive to the
details of the quasiparticle spectrum, we take the simplest spectrum, which
has the tetragonal symmetry and four nodes. In a bulk $d$-wave superconductor,
$E({\bf p})=\sqrt{\xi _p^2 +\Delta^2_\infty \sin^2(2\alpha)  }$.
Here $\xi _p=\epsilon (p)-E_F$.
We consider an isotropic Fermi surface around the $c$ axis and assume that the
vortex is axisymmetric. The magnetic field
is along the symmetry axis which is  the $z$ axis of the
coordinate frame with the positive direction along the vortex circulation
$\hat {\bf z}={\rm sgn}\, (e) {\bf B}/B$;
$\alpha$ is the asimuthal angle of ${\bf
p}$ counted from one of the 4 nodes in the gap ($\alpha=0$).

{\it Quasiparticles bound to the vortex}. The low-energy spectrum of
Caroli-de-Gennes-Matricon quasiparticles around a vortex contains an
anomalous branch.
Excitations on this branch are characterized by two canonically conjugated
variables, the angle
$\alpha$ and the angular momentum $Q$. In the quasiclassical approximation the
variables $\alpha$ and $Q$ are commuting, and the  spectrum is
$E(Q,\alpha)= -\omega_0(\alpha) Q $.
For a $d$-wave superconductor close to the gap nodes
$|\sin (2\alpha)|\ll 1$, the distance between the energy levels, i.e., the
minigap is
\cite{d-waveVortex,KopninVolovik}
\begin{equation}
\omega_0(\alpha)=\frac{2\Delta _\infty ^2\sin^2(2\alpha)}{v_\perp p_\perp}
\ln \frac{1}{\vert \sin(2\alpha)\vert} .\label{omega0}
\end{equation}
Here $p_\perp$, $v_\perp$ are the projections of the Fermi momentum and
velocity on the $x-y$ plane.
According to Eq. (\ref{omega0}), the minigap $\omega _0$ as a function of
$\alpha$ has nodes of a higher order than the gap itself. We shall show that
the strong singularity in $\omega _0(\alpha)$ results in a nonvanishing
dissipation even in the superclean limit $\Delta ^2_\infty \tau /E_F \gg 1$ when
the quasiclassical minigap $\omega _0(\alpha)$ is well resolved.

{\it Kinetic equation} for the distribution function $f(t,\alpha,Q)$
for a system  of fermions
characterized by canonically conjugated variables has a
conventional form \cite{Stone}:
\begin{equation}
{\partial f \over \partial t} - {\partial f\over \partial\alpha} {\partial
E\over \partial Q} +{\partial E\over \partial\alpha} {\partial
f\over \partial Q}= - { f-f_0\over \tau_0} .\label{KE}
\end{equation}
Here $\partial E/\partial Q=-\omega _0(\alpha)$; $f_0$ is the equilibrium
Fermi distribution;  the collision integral is written in a
$\tau$-approximation. A rigourous consideration shows that the
collision integral for localized excitations does not have singularities
as a function of $\alpha $, thus it can be considered as a constant on
order of the
relaxation time $\tau$ in the normal state.

If the  vortex moves with a constant velocity ${\bf u}=u(\hat {\bf x}\cos \phi
+ \hat {\bf y}\sin\phi)$ with respect to the lattice, the Doppler shift of
the energy
$E\rightarrow E(Q,\alpha)-{\bf p}{\bf u}$ produces the ``driving force''
$(\partial E/ \partial\alpha)=up_\perp \sin (\alpha-\phi)$ acting on
quasiparticles.  The third
term in the lhs of kinetic equation (\ref{KE}) contains this force multiplied
by $\partial  f_0/\partial Q=-\omega _0(\alpha)(d f_0/d E)$. Introducing the
longitudinal
and transverse response to this perturbation and taking into account that, for
the tetragonal symmetry, the response does not depend on the direction of the
vortex motion with respect to the crystal lattice, one obtains
\begin{equation}
\delta f =-up_\perp \frac{df_0}{dE}[\gamma _H\cos (\alpha -\phi )
+\gamma _O\sin (\alpha -\phi )]
\end{equation}
where $\gamma _O(\alpha )$ and $\gamma _H(\alpha )$ satisfy two
coupled first-order differential equations
\begin{eqnarray}
\frac{\partial \gamma _O}{\partial \alpha} -\gamma _H
-U(\alpha )\gamma _O +1 =0 ,\nonumber \\
\frac{\partial \gamma _H}{\partial \alpha} +\gamma _O
-U(\alpha )\gamma _H =0 \label{eq-gamma}
\end{eqnarray}
where $U(\alpha)=\left[\omega _0(\alpha)\tau \right]^{-1} $.
These two real functions are combined into one complex function
$\gamma _O={\rm Im}\, W(\alpha )$, $\gamma _H = {\rm Re}\, W(\alpha )$,
which obeys the equation
\begin{equation}
(\partial W/\partial \alpha ) - (U+i)W +i =0.  \label{eq-W}
\end{equation}

{\it The force acting on the vortex} from the environment where it moves  is
the momentum transferred from the excitations:
\begin{equation}
{\bf F}_{\rm env} = - \int \frac{dp_z}{2\pi} \frac{dQ\, d\alpha}{2\pi}
\frac{\partial {\bf p}}{\partial t}\delta f
\end{equation}
Using $\partial {\bf p}/\partial t =-\nabla E=[\hat {\bf z}\times {\bf p}]
(\partial E/\partial Q)$ we get
\begin{eqnarray}
{\bf F}_{\rm env}
=-\int \frac{dp_z}{2\pi} \frac{dQ\, d\alpha}{2\pi}
[\hat {\bf z}\times {\bf p}_\perp ]\frac{\partial E}{\partial Q}\delta f
\nonumber \\ =
\pi n\left(\left<\gamma _H\right>_F
[\hat {\bf z}\times{\bf u}] - \left<\gamma _O\right>_F {\bf u}\right) .
\end{eqnarray}
Here $n$ is the electron density; $\left<\cdots \right>_F$ is the
average over the whole Fermi surface.
The force ${\bf F}_{\rm env}$ should be balanced by the Lorentz force
${\bf F}_L=\pi n[{\bf v}_s\times \hat {\bf z}]$. The so called mutual
friction parameters $\left<\gamma _O\right>_F$ and $\left<\gamma _H\right>_F$
are thus coupled to the
flux-flow longitudinal and Hall conductivities. At low
temperatures,
\begin{equation}
\left<\gamma _O\right>_F =(B\sigma _O/n |e| c) ,~\left<\gamma _H\right>_F
=(B\sigma _H/n e c) .
\end{equation}
If the Fermi surface has electron-like and hole-like pockets,
the conductivities at low temperatures are \cite{KL}
\begin{eqnarray}
\sigma _O=
\frac{|e|c}{B}\left[ n_e\left<\gamma _O\right>_{F,e}+
n_h \left<\gamma _O\right>_{F,h}\right] ,\label{sigmaO} \\
\sigma _H =
\frac{ec}{B}\left[ n_e\left<\gamma _H\right>_{F,e}-
n_h \left<\gamma _H\right>_{F,h}\right] . \label{sigmaH}
\end{eqnarray}
where $\left<\ldots \right>_{F,e}$ and $\left<\ldots \right>_{F,h}$
are the averages over the electron-like and hole-like parts of the Fermi
surface, respectively; $n_e$ and $n_h$ are the numbers of electrons and
holes.

{\it Solution of the kinetic equation} (\ref{eq-W}) is
\begin{equation}
W=e^{[i\alpha +F(\alpha )]}\left( C-i\int _0^\alpha e^{-[i \alpha ^\prime
+F(\alpha ^\prime)]}\, d\alpha ^\prime \right) \label{W-sol}
\end{equation}
where
$
 F(\alpha )=\int _0^\alpha U(\alpha ^\prime )\, d\alpha ^\prime
$.
In the moderately clean limit, $\Delta_0^2\tau /E_F\ll 1$,
the potential $U(\alpha )$
is always large, and we obtain the local solution as in
an $s$-wave superconductor \cite{KL}
\begin{equation}
\gamma _0(\alpha )=\omega _0(\alpha )\tau ,~
\gamma _H(\alpha )=[\omega _0(\alpha )\tau ]^2 .
\end{equation}
The conductivities are
\begin{equation}
 \sigma _O\sim \frac{nec}{B}\, \frac{\Delta _\infty ^2\tau}{E_F}\,
\ln \frac{T_c}{T}
\end{equation}
and $\sigma _H/\sigma _O \sim (\Delta _\infty ^2\tau/E_F)\ln (T_c/T) $
which are similar to the results for an $s$-wave superconductor
\cite{KopKr,KL,LO1}.

In the superclean limit, $\Delta_0^2\tau /E_F\gg 1$, the situation is more
interesting. The potential $U(\alpha)$ is small almost everywhere
except for vicinities of the gap nodes where it diverges:  $U(\alpha)\sim
1/[\Gamma (\delta\alpha )^2]$ at $\vert \delta \alpha\vert \ll 1$.Here
$\Gamma \sim \Delta_0^2\tau /E_F$.
However there is a physical infrared cutoff
$\alpha_{min}$ for the angle below which the divergent potential $U(\alpha)$
saturates.  For example,
if the vortices are separated by a finite distance $R\sim \xi \sqrt{H_{c2}/B}$
with $B\ll H_{c2}$, i.e., $R\gg\xi $,
the Doppler shift of the energy ${\bf p}_F{\bf v}_s\sim T_c(\xi /R)$
(where ${\bf v}_s$ is the vortex-induced superflow velocity)
affects the spectrum Eq. (\ref{omega0}) when the excitation energy
becomes comparable with the gap $\Delta _0|\sin (2\alpha )|$. This
determines the cutoff $\alpha_{min}=\sqrt{B/H_{c2}}$
\cite{d-waveVortex,KopninVolovik}. In addition, a finite temperature introduces
a cutoff $\alpha_{min}=T/T_c$ since an excitation with a smaller $\alpha$ is
already above the gap $\Delta _0|\sin (2\alpha )|$.
The relaxation rate $1/\tau$ also provides a cutoff in the form
$\alpha_{min}=1/\tau T_c$. Since the parameter $1/\tau T_c$ is
much smaller than all the other
cutoff parameters in our case, we can write in general
\begin{equation}
\alpha_{min}={\rm max} \left({T/T_c},\sqrt{B/H_{c2}} \right) .\label{cutoff}
\end{equation}

Assuming that $U(\alpha <  \alpha_{min})=1/(\Gamma \alpha_{min}^2)$
one can insert the truncated potential  into Eq. (\ref{W-sol}).
Since the potential $U(\alpha)$ is concentrated near the nodes one can
neglect the potential almost in the whole range of $\alpha$
outside a small vicinity of the nodes.   Eq. (\ref{W-sol}) has the form
$W=1 + Ae^{i\alpha}$
where the complex constant $A$ can be found by matching with the
solution in the vicinty of the node, $\alpha \ll 1$, i.e.,
$W(\alpha) \propto e^{F(\alpha )} $.
This provides the boundary condition
$W(+0) = e^{2\lambda}  W(-0) $ across the node at $\alpha=0$. Here
\begin{equation}
2\lambda=F(+0)-F(-0)
\sim 1/(\Gamma \alpha_{min}) . \label{lambda}
\end{equation}

The response function has the same tetragonal symmetry as the
underlying system,
$W(\alpha +\pi/2) = W(\alpha)$. Together with the above boundary condition,
this gives
\begin{equation}
W({\alpha}) = 1   - {1 -e^{2\lambda} \over 1 -ie^{2\lambda}} e^{i\alpha}
~, ~0<\alpha<\pi/2 ,
\end{equation}
which is to be periodically continued to the rest of angles with the
period $\pi/2$. We have
\begin{equation}
\left<\gamma _H\right>
=1-\frac{4}{\pi}\frac{1 }{ 1+\coth ^2\lambda},~
\left<\gamma _O\right> =\frac{4}{\pi}
\frac{\coth \lambda }{1+\coth ^2\lambda}. \label{gammas}
\end{equation}
Here $\left<\cdots \right>$ is the average over the azimuthal angle $\alpha$.

{\it Magnetic field and temperature dependence}. In the low-field limit
$B/H_{c2}\ll (T/T_c)^2$ we have
$\lambda =T_c/(T\Gamma )$.
At not very low temperatures $1/\Gamma \ll T/T_c \ll 1$, the parameters
in Eq. (\ref{gammas}) are
\begin{equation}
\left<\gamma _O \right>\sim T_c/(T\Gamma ),~
1- \left<\gamma _H\right>\sim  [ T_c/(T\Gamma)]^2 .
\end{equation}
For $T\sim T_c$, these expressions transform into usual solutions for a
superclean
$s$-wave superconductor \cite{KopKr,KL}. For very low temperatures
$T/T_c \ll 1/\Gamma$, the parameters become independent of the scattering
time
\begin{equation}
\left<\gamma _O \right> =2/\pi ,~
\left<\gamma _H\right> =1- (2/\pi ) . \label{univers}
\end{equation}

For fields $B/H_{c2}\gg (T/T_c)^2$, we have
$ \lambda= ( 1/\Gamma )\sqrt{H_{c2}/B}$.
The asymptotics for $1/\Gamma^2 \ll B/H_{c2}$ are
\begin{equation}
\left<\gamma _O \right> \sim  (1/\Gamma ) \sqrt{H_{c2}/B }   ~,
1- \left<\gamma _H\right>\sim H_{c2}/(\Gamma ^2B) . \label{Bdepend}
\end{equation}
Eq. (\ref{Bdepend}) suggests the magnetic-field dependences of the
flux-flow conductivity in the form $\sigma _O\propto B^{-3/2}$.
In the limit $B/H_{c2}\ll 1/\Gamma^2 $  we recover Eq. (\ref{univers}).

{\it Universal superclean limit}. For $\Gamma \gg 1$ and for low temperatures
and fields such that
\begin{equation}
T/T_c~{\rm and} ~\sqrt{B/H_{c2}} \ll 1/\Gamma ,\label{condition}
\end{equation}
the longitudinal and transverse conductivities approach finite universal values
independent of the scattering mean free time $\tau$:
\begin{equation}
\sigma _O=\frac{(n_e+n_h)|e|c}{B}\left<\gamma _O\right>; ~
\sigma _H=\frac{(n_e-n_h)ec}{B}\left<\gamma _H\right> \label{sigma/univ}
\end{equation}
where $\left<\gamma _O\right>$ and $\left<\gamma _H\right>$
are given by Eq. (\ref{univers}).
The scattering rate affects only the range of temperatures and fields,
Eq. (\ref{condition}),
where this asymptotic behavior is established.
With lowering the temperature for  fixed $\Gamma \gg 1$ and $B$,
one expects an increase in $\sigma _O$ and, finally, a crossover
to the regime where both longitudinal and
Hall conductivities have universal values independent of the mean free path.
Together with the conductivities, the Hall angle reaches its universal
value determined only by the numbers of electron and holes: $\tan \theta _H=
(\pi /2-1)(n_e-n_h)/(n_e+n_h)$. The universal value of $|\tan \theta _H |$
is expected to be a local minimum at low temperatures due to an increasing
$\sigma _O$.

Experimentally, the  sample purity needed for observation of
the universal  behavior of Eq. (\ref{condition}) is rather high;
it requires $\ell /\xi _0 \gg 10$ to 100 depending on the ratio $E_F/T_c$
for the particular high-$T_c$ superconductor. Nevertheless,
there are indications that such a regime can be realized in practice:
the superclean limit was nearly approached in
Ref. \onlinecite{Ong} with $\Gamma \sim 1$ for a presumably $d$-wave
60 K YBaCuO compound \cite{Ott}.

{\it Landau damping}. To understand the reason for a finite dissipation in the
limit of an infinite relaxation time
$\tau=\infty$,  let us consider the $\omega$-dependence of the response function
$W$. It is obtained by the substitution $i/\tau \rightarrow \omega$. Therefore,
$\lambda \to -i\pi\omega /4E_0$ where
\begin{equation}
E_0=\left[ \frac{1}{2\pi}\int_0^{2\pi}
\frac{d\alpha}{\omega _0(\alpha)} \right]^{-1} . \label{E0}
\end{equation}
We have
\[
\left<\gamma _H(\omega )\right> +i\left<\gamma _O(\omega )\right> =
1 + \frac{2(1+i)}{\pi} \frac{1-\exp (-i\pi \omega /2E_0)}
{1 -i\exp (-i\pi \omega /2E_0)} .
\]
This response function has
poles  at $\omega_k= (4k+1)E_0$ where $k$ is an integer.
Note that $E_0$ in Eq. (\ref{E0}) is the true interlevel distance in the
spectrum of bound states. It can be obtained by  quantization of the azimuthal
motion using the Bohr-Zommerfeld rule
\cite{KopninVolovik}: $\oint Q(\alpha)d\alpha =2\pi (m +
\gamma)$, where $m$ is an integer and $\gamma\sim 1$. Since
$Q(\alpha)=-E/\omega_0(\alpha)$,  the quantum-mechanical spectrum is
$E=-(m+\gamma)E_0$, assuming that the integral in Eq. (\ref{E0})
converges due to a finite magnetic field.  The poles correspond to
eigenmodes in the quasiparticle distribution when the frequency matches the
resonance transition between the levels with different $m$. In $s$-wave
superconductors,  the selection rule for transitions caused
by motion of an axisymmetric vortex is
$\Delta m=\pm 1$ with the resonance at $\omega =E_0$. In $d$-wave
superconductors other harmonics are mixed due to the tetragonal symmetry
giving rise to  resonances at $\omega =\omega _k$.

When $\alpha_{\rm min}\rightarrow 0$, thus $E_0 \to 0$, the
poles densly fill the $\omega$-axis. This leads to a finite density of states
and thus to a finite Ohmic resistivity at $\omega=0$: the parameter
\[
 \left<\gamma _O\right> = \frac{2}{\pi}\, {\rm lim}_{\, E_0 \to 0}
\, {\rm Im}\, \left[ \frac{1+i}{\Omega} \int_0^\Omega d\omega \frac{1
-e^{-i\pi \omega /2E_0}}{1 -ie^{-i\pi \omega /2E_0}}\right]
\]
goes to $2/\pi $ as $\Omega \to 0$.
The attenuation in the zero-frequency limit is a special type of the Landau
damping of vortex motion due to the nodes in the minigap.

The true interlevel distance $E_0$ introduces a new energy scale which does
not exist for superconductors without gap nodes. For conventional
superconductors,  there is a single characteristic value, the minigap
$\omega _0$, which marks
a crossover in the relaxation rate $1/\tau$ between two regimes:
a dissipative viscous flux flow for $\omega _0\tau \ll 1$, and a
dissipationless vortex motion for $\omega _0\tau \gg 1$.
For superconductors with gap nodes, there are two critical values: the
average quasiclassical interlevel distance $\left<\omega_0\right>$ and the true
interlevel distance $E_0$ which can be less than $\left<\omega_0\right>$.
Indeed, in the superclean limit $\left<\omega _0\right>\tau \gg 1$,
the true minigap is $E_0 \sim \left<\omega _0\right> \alpha _{min}$;
near the nodes, the quasiclassical levels approach each other on
much shorter distances which can only be resolved by quantization of the
angular motion.

The condition of Eq. (\ref{condition}) means that $E_0 \tau \ll 1$. A
finite density of states appears when the true quantum-mechanical interlevel
distance is smaller than the level width.
Therefore one has three regimes: (i) Moderately clean regime
$\left< \omega _0 \right>\tau \ll 1$ with a high dissipation, (ii)
intermediate universal regime $E_0 \tau \ll 1$ with a finite
dissipation at zero $\omega$, where the quasiclassical
levels are well resolved on average but the true minigap is
smaller than the scattering rate; and
(iii) the extreme superclean case $E_0 \tau \gg 1$ when the true
levels are well separated and thus there is no dissipation.

The regime with an universal flux-flow
conductivity is reminiscent of the asymptotic behavior of the current
response function in the Meissner state. In the flux-flow state, however,
the regime of the universal conductivity is reached at higher temperatures,
$T\ll T_c/\Gamma $. The reason is that the minigap in the
spectrum of low-energy excitation which are responsible for
the vortex dynamics, has nodes of a higher order than the gap itself.
Therefore, the effect of the nodes in the flux-flow regime is more pronounced.
The universal values of the conductivities correspond to an
effective scattering rate $1/\tau $ of the order of the characteristic
minigap $\omega _0$. This is also in parallel with the behavior of the
response function in the Meissner state, where the effective scattering rate is
of the order of the characteristic gap $\Delta _\infty$ \cite{Lee}.

{\it In conclusion}, we have calculated the flux-flow dynamics in $d$-wave
superconductors at low temperatures. In the moderately clean case,
$\Delta ^2_\infty \tau /E_F \ll 1$, the Ohmic and Hall components of the
flux-flow conductivity are similar  to those
for conventional superconductors. However, for superclean systems with
$\Delta ^2_\infty \tau /E_F \gg 1$, the magnetic field dependence is unusual.
Moreover, at low $B$ and $T$, both Ohmic and Hall components
approach universal values independent of relaxation. A finite
dissipation in the superclean limit is explained in terms of the Landau
damping on zero-frequency vortex modes which appear due to  minigap
nodes in the spectrum of bound states in the vortex core. Since $\Delta
_\infty/E_F$ is
not very small in high-$T_c$ compounds, the superclean regime is
accessible in those high purity crystals, where the scaling law for the specific
heat has been verified \cite{Junod}.

This work was supported by the Russian Foundation for Fundamental Research
grant No. 96-02-16072, by the RAS program
``Statistical Physics'', and partially by Swiss National Foundation
cooperation grant No. 7SUP J048531.

\end{document}